\documentstyle[psfig,twocolumn,prb,aps]{revtex}

\begin{document}
\draft

\wideabs{

\title{\bf KINKS\ IN\ A\ PERIODICALLY\ MODULATED\ DISORDERED\ SYSTEM}

\author{Eva Majern\'{\i}kov\'a${}^{\dag,\dag\dag}$ and Jaroslav
Riedel$^{\dag}$}
\address{${}^{\dag}$Department of Theoretical Physics, Palack\'y University,
T\v r. 17. Listopadu 50, CZ-77207 Olomouc, Czech Republic\\
${}^{\dag\dag}$Institute of Physics, Slovak Academy of
Sciences, D\'ubravsk\'a cesta 9, SK-84228 Bratislava, Slovak Republic}
\author{Boris A. Malomed}
\address{Department of Interdisciplinary Studies, Faculty of Engineering,
Tel Aviv University, Tel Aviv 69978, Israel}

\date{\today}

\maketitle

%%%%%%%%%%%%%%%%%%%%%%%%%%%%%%%%%%%%%%%%%%%%%%%%%%%%%%%%%%%%%%%%%%
\begin{abstract}
We consider a dc-driven damped sine-Gordon model with a small
nonlinear spatial-disorder term, onto which a sinusoidal
modulation is superimposed. It describes, e.g., a weakly  disordered system
with a regular grain structure. We demonstrate that, at the
second order of the perturbation theory (with respect to the weak spatial
disorder), the periodically modulated disorder gives rise to an
effective periodic potential. Dynamics of a kink moving in this potential is
studied in the overdamped limit, using the adiabatic
approximation, the main objective being to consider depinning of a
trapped kink. The analytical
results are compared with direct dynamical simulations of the underlying
model, as well as with numerical results using the collective-coordinate
approach but without the mean-field approximation. It is found that a
critical force for the depinning of a kink trapped by the periodically
modulated weak spatial disorder is much larger than that predicted by the
mean-field approximation.
\end{abstract}

\pacs{PACS: 05.45.Y, 61.43.Bn, 71.55.Jv}
}
\narrowtext

\section{Introduction}

The interplay of nonlinearity and disorder is a well-known topic in
theoretical and experimental studies of solitons
\cite{Bishop:1989:DN,Konotop:1994:NRW,Malomed:1989:PRB,Mingaleev:1999:PRB}.
A paradigm one-dimensional model which has
numerous physical applications and, in particular, allows one to consider
interaction of topological solitons (kinks) with a random spatial
inhomogeneity is a perturbed sine-Gordon (SG) equation. In the simplest
case, the corresponding model takes a form
\cite{Malomed:1989:PRB,Gredeskul:1992:PRA},
\begin{equation}
\Phi _{tt}-\Phi _{xx}+\left[ 1+\epsilon (x)\right] \sin \Phi =-(\Gamma
/8)\Phi _{t}-f,  \label{SG}
\end{equation}
where $\Phi $ is a dynamical order parameter, $\Gamma $ is a dissipation
constant, $f={\rm const}$ represents a driving force (external dc field),
and the disorder function $\epsilon (x)$ is determined by its white-noise
Gaussian correlations, 
\begin{equation}
\langle \epsilon (x)\rangle =0,\,\,
\langle \epsilon (x)\epsilon (x^{^{\prime
}})\rangle =\eta ^{2}\delta (x-x^{^{\prime }}),  \label{wnoise}
\end{equation}
where $\eta$ is the disorder amplitude. In this work, the perturbation
coefficients $\eta ^{2}$ and $f$ will be treated as small parameters, while $%
\Gamma $ is not necessarily small. Note, that similar disordered models
can be defined not only in the continuum form, but also on aperiodic
lattices \cite{Dominguez:1995:PRE}.

In some systems, the spatial disorder can be subject to a superimposed
regular (periodic) modulation. For instance, this may be the case if the
underlying slightly disordered physical medium has a regular grain
structure. Then, the model (\ref{SG}) is modified as follows, in the
simplest case when the superimposed modulation is sinusoidal,
\begin{equation}
\Phi _{tt}-\Phi _{xx}+\left[ 1+\epsilon (x) \sin (kx)\right] \sin \Phi
=-(\Gamma /8)\Phi _{t}-f,  \label{pert}
\end{equation}
$2\pi /k$ being the modulation period.

As concerns the form of the model (\ref{pert}), it is relevant to
mention that, for any finite $k$, one can replace the modulation
function $\sin (kx)$ by $\cos(kx)$, shifting $x \to x+\pi/2k $. If the
system is very long and $\epsilon(x)$ obeys the Gaussian correlations
(\ref{wnoise}), the modulation functions are tantamount to each other,
except the case $k=0$ (or $|k|\, ^{<}_{\sim}\, 1/l$, if the system has
a finite length $l$). Actually, for $k=0$ one comes back to the model
(\ref{SG}) with the unmodulated spatial disorder.

Dynamics of topological nonlinear excitations in the form of kinks in
the model (\ref{pert}) is a subject of the present work.
In the lowest-order approximation,
the random perturbation does not induce any effective potential for the
kink. However,
we demonstrate in section 2 that an effective periodic potential is
generated at the second order of the perturbation theory. It is interesting
that the amplitude of the effective potential vanishes at $k^{2}=1/2$.

 In section 3 we study motion of the driven kink in this effective potential.
The corresponding equation of motion is derived in the adiabatic
approximation (treating the kink as a quasiparticle). This approximation
is a version of the well-known collective-coordinate technique, which was
applied to the (weakly) disordered models in various contexts (see, e.g.,
 Refs. \cite{Malomed:1989:PRB,Dominguez:1995:PRE}), allowing one
not only to derive an effective equation of motion, but also to study
dynamical statistical characteristics of the kink
\cite{Dominguez:1995:PRE}.
We focus on the case
when the {\it overdamped}  approximation applies to the description of
the kink motion (i.e., the inertia term in the corresponding equation of
motion is neglected). In its direct form, the latter approximation
assumes that the dissipation is stronger than the disorder, or, in terms
of Eq. (\ref{SG}) $\Gamma \gg \eta $, in
the generic case with $k\sim 1$. However, in this work we actually
focus on a transition between the pinning and free-motion dynamical
regimes (i.e., depinning of the trapped kink) with the increase of the
driving force $f$. It is quite reasonable to assume that the inertia
plays a negligible role in the depinning process, hence the overdamped
approximation can be applied in a broad parametric region. Note that
depinning of kinks trapped by the unmodulated spatial disorder in the
model (\ref{SG}) was earlier studied in detail in Ref.
\cite{Malomed:1989:PRB}.

Section 4 displays results of numerical simulations of the
model, which consists of two parts: direct dynamical simulations, and
numerical analysis of the collective-coordinate description without
using the mean-field approximation.  The latter approach yields
accurate results for the critical (minimum) force $f_{\rm cr}$ necessary
to depin the trapped kink. The critical force turns out to be much
larger than that predicted by the mean-field approximation. In
particular, $f_{\rm cr}$ is predicted by the mean-field approximation to
scale $\sim \eta^2$ (see Eq. (\ref{wnoise})), while the actually found
critical force scales $\sim\eta$, the dependence on the modulation wave
number $k$ (see Eq. (\ref{pert})) also being quite different.

\section{Sine-Gordon kink in a periodically modulated random potential}

Equation ({\ref{pert}}) can be solved by iterations, assuming a small
amplitude of the noise. The unperturbed kink solution is
\begin{equation}
\Phi_0 (x-\xi)=4\tan^{-1}\exp(x-\xi)\,,\label{kink}
\end{equation}
$\xi $ being the coordinate of the kink's center.

A first-order correction to the kink waveform,
$\phi _{1}=\Phi (x)-\Phi _{0}$, satisfies the equation
\begin{equation}
(\phi _{1})_{tt}-(\phi _{1})_{xx}+\cos \Phi _{0} \phi _{1}=-\epsilon
(x) \sin (kx) \sin \Phi _{0}.  \label{linear}
\end{equation}
To solve Eq. (\ref{linear}), we need its Green's function, which is
defined in a standard way by means of the equation,

\begin{equation}
\left\{ \frac{\partial ^{2}}{\partial t^{2}}-\frac{\partial ^{2}}{\partial
x^{2}}+V^{\prime \prime }[\Phi _{0}(x)]\right\} G(x,x^{\prime },\tau
)=\delta (x-x^{\prime})\delta (\tau ),
\label{Green}
\end{equation}
where $ \tau =t-t^{\prime}$ and $V(\Phi )=1-\cos \Phi $ is the SG potential.
To simplify the formulas, we temporarily set $\xi=0$.

 The necessary Green's function has been found
in an explicit form by Flesch and Trullinger \cite{Flesch:1987:JMP}
(see Eq. (3.29)):\\
$G^{\rm SG}(x,x^{\prime},\tau )$\\
\begin{eqnarray}
 =[\theta (\tau -|z|)/2]\left\{
J_{0}(s)+\beta _{2}[sJ_{1}(s)-2\tau
\Lambda _{1}(w,s)]\right.\nonumber\\
\left. -\beta _{3}{\rm sgn}(z)[-(\tau +|z|)J_{0}(s)+2\tau \Lambda
_{0}(w,s)]\right\}.  \label{Greensg}
\end{eqnarray}
Here, $\ z =x-x^{\prime}$, $ s\equiv \sqrt {\tau^2-z^2}$, $w\equiv
\tau-|z|$, $\theta (\tau)$ is the Heaviside's step function, and
$\beta _{2}\equiv \tanh (x)\tanh (x^{\prime })-1,\ \beta _{3}\equiv
\tanh (x^{\prime})-\tanh (x)$. Further, $J_0(s)$ and $J_{1}(s)$ are the Bessel
functions and  $\Lambda _{0}(w,s)$
and $\Lambda _{1}(w,s)$ are the modified Lommel functions defined as 
\begin{equation}
\Lambda _{n}(w,s)=\sum\limits_{m=0}^{\infty }\left( \frac{w}{s}\right)
^{2m+n}J_{2m+n}(s)\,.  \label{lommel} 
\end{equation}
%%%%%%%%%%%%%%%%%%%%%%%%%%%%%%%%%%%%%%%%%%%
\begin{figure}[h]
\centerline{\hbox{
\psfig{figure=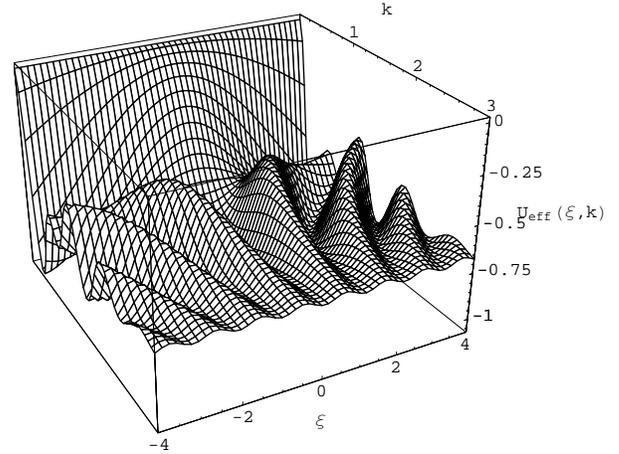,width=80mm,angle=0}}}
\caption{The effective potential $U_{{\rm eff}}(k,\xi )$.}
\end{figure}
%%%%%%%%%%%%%%%%%%%%%%%%%%%%%%%%%%%%%%%%%%%

Then,
$\phi _{1}$ can be obtained in the form
\begin{eqnarray}
\phi _{1}(k,x)
=-\int\limits_{-\infty }^{\infty }dx^{^{\prime }}\zeta
(x^{\prime })\sin (\phi _{0}(x^{\prime}-\xi ))
\nonumber\\
\times\sin (kx^{\prime})G(x-x^{\prime})\,,  \label{corr}
\end{eqnarray}
where we have restored the kink's coordinate $\xi$.

The perturbation on the right-hand side of Eq.(\ref{linear}) gives rise, in
a straightforward way \cite{Kivshar:1989:RMP}, to an effective potential for the
 kink treated as a quasiparticle:
\begin{equation}
U_{{\rm eff}}(k,\xi )=\int\limits_{-\infty }^{\infty }dx\,\epsilon (x)\sin
(kx)\phi _{1}(k,x)\sin \Phi _{0}(x-\xi ).  \label{effpot}
\end{equation}
 Then, inserting the expression (\ref
{corr}) for $\phi _{1}(k,x)$ into Eq. (\ref{effpot}), one obtains 
\begin{eqnarray}
U_{{\rm eff}}(k,\xi ) =-\int\limits_{-\infty }^{\infty
}dx\int\limits_{-\infty }^{\infty }dx^{^{\prime }}\epsilon (x)\epsilon
(x^{^{\prime }})\sin \Phi _{0}(x-\xi )
\nonumber \\
\times\sin \Phi _{0}(x^{^{\prime }}-\xi)\sin (kx)\sin (kx^{^{\prime }})
G(x-x^{^{\prime }}).  \label{Effpot2}
\end{eqnarray}

The potential (\ref{Effpot2}) can be averaged with regard to (\ref{wnoise}).
Using the elementary expressions for the SG kink, $\cos \Phi _{0}(x)=1-2{\rm %
sech}^{2}x\ $ and $\ \sin \Phi _{0}=2\sinh x {\rm sech}^{2}x$,$\ $
yields the {\it average potential} 
%%%%%%%%%%%%%%%%%%%%%%%%%%%%%%%%%%%%%%%%%%%
\begin{figure}[hbt]
\centerline{\hbox{
\psfig{figure=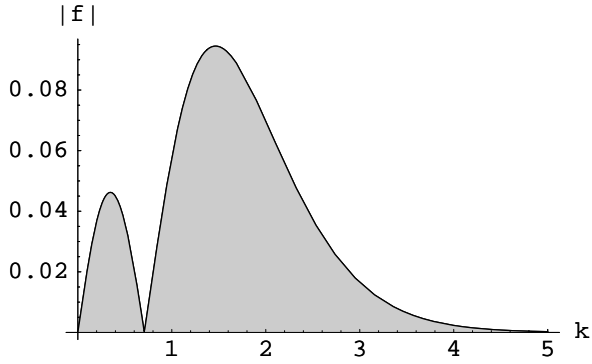,width=80mm,angle=0}}}
\caption{The phase diagram for $\eta =1$. The regions of the free-motion ($%
|A/B|<1$) and pinning ($|A/B|>1$) regimes are, respectively, white and grey.
At the boundary between them, $|A/B|=1$.}
\end{figure}
%%%%%%%%%%%%%%%%%%%%%%%%%%%%%%%%%%%%%%%%%%%
\begin{eqnarray}
\langle U_{{\rm eff}}(k,\xi )\rangle & = &-2\eta ^{2}
\int\limits_{-\infty }^{\infty }dx\frac{\sinh ^{2}(x-\xi )}
{\cosh ^{4}(x-\xi )}\sin^{2}(kx)\nonumber\\
& = & -\frac{2\eta ^{2}}{3}\left[ 1-\pi k(1-2k^{2})\frac{\cos (2k\xi )}{%
\sinh (\pi k)}\right] \,.  \label{avpot}
\end{eqnarray}
 
Evidently, the average potential $\langle U_{{\rm eff}}(k,\xi )\rangle $ (%
\ref{avpot}) is periodic. It is plotted in Fig. 1 as a function of $\xi $
and $k$.

Minima of the potential (\ref{avpot}) are at the points $\xi _{{\rm min}%
}=\left( \pi /2k\right) (2n+1),\ n=0,1,...$, for $k^{2}<1/2,$ and $\xi _{%
{\rm min}}=\left( \pi /k\right) n$ for $k^{2}>1/2$. Close to the minima, $%
\xi =\xi _{{\rm min}}+\delta ,\ k\delta <<1$, the potential gives rise to
small nearly harmonic oscillations with the frequency $\Omega ^{2}=4\eta
^{2}\pi k^{2}|2k^{2}-1|/\left( 3\sinh \pi k\right) $.

\section{Driven motion of the kink in the effective potential: adiabatic
approximation}

In order to investigate uniformly driven damped motion of the kink in the
effective periodically modulated random potential (\ref{avpot}), we derive,
in the framework of the standard adiabatic (quasi-particle) approximation 
\cite{Kivshar:1989:RMP}, an equation of motion for the kink's center,
\begin{equation}
8\frac{d^{2}\xi }{dt^{2}}+\Gamma \frac{d\xi }{dt}=2\pi f+\frac{4\eta ^{2}\pi
k^{2}(1-2k^{2})}{3\sinh (\pi k)}\sin (2k\xi )  \label{11}
\end{equation}
(here, the ``nonrelativistic'' approximation, $\left( d\xi /dt\right)
^{2}\ll 1$, is additionally adopted; note that the kink's mass in the
present notation is $m=8$ \cite{Kivshar:1989:RMP}).

As it was explained above, we will chiefly be interested in the depinning
threshold, when the inertia term in Eq. (\ref{11}) may be neglected.
Then, the equation of motion simplifies to
\begin{equation}
\frac{d\xi }{dt}=\frac{2\pi f}{\Gamma }+\frac{4\eta ^{2}\pi k^{2}(1-2k^{2})}{%
3\Gamma \sinh (\pi k)}\sin (2k\xi )\equiv B+A\sin (2k\xi ),  \label{12}
\end{equation}
where $A\equiv 4\eta ^{2}\pi k^{2}(1-2k^{2})/\left[ 3\Gamma \sinh (\pi k)%
\right] $ and $B\equiv 2\pi f/\Gamma $. Comparing Eqs. (\ref{11}) and (\ref
{12}), one can verify that the inertia term is indeed negligible provided
that $\Gamma \gg |A| k$. In the general case ($k\sim 1$), this simply means $%
\Gamma \gg \eta $, i.e., the dissipative constant must be essentially larger
than the amplitude of the spatial disorder.

A solution to Eq. (\ref{12}) is obvious. One has to distinguish the cases
(i) $|A|=|B|$, (ii) $|A|<|B|$, and (iii) $|A|>|B|$. The driving and
potential terms dominate, respectively, in the regions (ii) and (iii), while
(i) represents a border between them, characterized by the value 
\begin{equation}
f_{{\rm b}}=(2k^{2}\eta ^{2}/3)|1-2k^{2}|/\sinh (\pi k)  \label{border}
\end{equation}
of the driving field. The corresponding phase diagram is plotted, in terms
of the underlying parameters $k$ and $f$, in Fig. 2, where the boundary
between the regions (iii) (grey) and (ii) (white) corresponds to the
critical field (\ref{border}). Basic features of these regimes, that can be
deduced in an analytical form from Eq. (\ref{12}), are summarized below.
%%%%%%%%%%%%%%%%%%%%%%%%%%%%%%%%%%%%%%%%%%%
\begin{figure}[ht]
\centerline{\hbox{
\psfig{figure=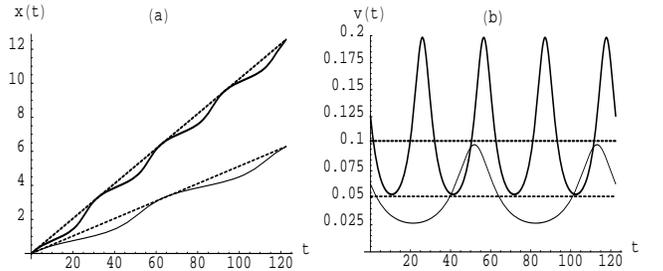,width=90mm,height=40mm,angle=0}}}
\caption{Free (progressive) motion of the kink, shown in terms of $x(t)$ (a)
and velocity $v(t)$ (b). Here and in the figures below 
the results are displayed for $\Gamma =10$ and $\Gamma =5$ by the thin and
thick lines. The other parameters are $\eta =1$, $k=1$, and $%
f=0.1$.}
\end{figure}
%%%%%%%%%%%%%%%%%%%%%%%%%%%%%%%%%%%%%%%%%%%
(i) The border regime, $\left| A/B\right| =1$: Eq. (\ref{12}) gives rise to 
{\it semi-stable} fixed (stationary) points, $\xi _{{\rm s}}=(4n-1)(\pi /4k)$%
,$\ n=0,\pm 1,...$, which seem like attractors and repellers to the left and
right of them, respectively. In this case, an analytical solution to Eq. (%
\ref{12}) takes an implicit form, 
\begin{equation}
-\frac{\cos (2k\xi )}{2kA(1+\sin (2k\xi ))}=t-t_{0},  \label{13}
\end{equation}
$t_{0}$ being an initial moment.

(ii) The free (progressive) motion regime, $\left| A/B\right| <1$. In this
case, Eq. (\ref{12}) does not have stationary solutions. A time-dependent
solution is\\
$\tan \left( k\xi (t)\right)$
\begin{equation}
 =-A/B+\sqrt{1-A^{2}/B^{2}}\tan \left[ k\sqrt{%
B^{2}-A^{2}}(t-t_{0})\right] .
\label{14}
\end{equation}

A typical example of this solution is displayed in Fig. 3. It represents a
superposition of a linear trajectory and 
%%%%%%%%%%%%%%%%%%%%%%%%%%%%%%%%%%%%%%%%%%%
\begin{figure}[t]
\centerline{\hbox{
\psfig{figure=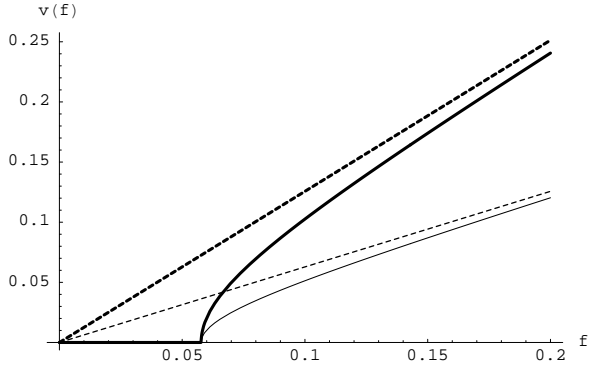,width=80mm,angle=0}}}
\caption{The average velocity $\overline{v}$ vs. the driving field for $k=1$, 
$\eta =1$. The phase transition occurs at $f_{b}=0.0577264$.}
\end{figure}
%%%%%%%%%%%%%%%%%%%%%%%%%%%%%%%%%%%%%%%%%%%
periodic oscillations. The linear
trajectory corresponds to the average velocity $\bar{v}\equiv
\lim\limits_{T\rightarrow \infty }\frac{1}{T}\left[ \xi (t_{0}+T)-\xi (t_{0})\right]
=\sqrt{B^{2}-A^{2}}$. In a neighborhood of the free-motion's border, at $%
f-f_{b}\,\rightarrow +0$, one has 
\begin{equation}
\bar{v}\approx \left( 2\pi /\Gamma \right) \sqrt{2f_{{\rm b}}\left( f-f_{%
{\rm b}}\right) }.  \label{14x}
\end{equation}
The average-velocity dependence as per Eq. (\ref{14x}) exhibits a typical
{\it mean-field} critical (cusp) behavior close to the border, $\bar{v}$ playing
the role of an order parameter (Figs. 4 and 5).

The time-periodic component of the velocity (see Fig. 3b) is characterized
by the frequency $\omega =2k\sqrt{B^{2}-A^{2}}$, whose critical behavior at $%
f-f_{b}\,\rightarrow +0$ is the same as that of the velocity (\ref{14x}).
The mean-square deviation of the trajectory from the linear one, $
\Delta
\equiv \lim\limits_{T\rightarrow \infty }
 \frac{1}{T}\int\limits_{0}^{T}dt(%
\xi (t)-\overline{v}t-\overline{\zeta })^{2} $, where $\overline{%
\zeta} \equiv \lim\limits_{T\rightarrow \infty}  \frac{1}{T}%
\int\limits_{0}^{T}dt(\xi (t)-\overline{v}t)$, exhibits too the
critical behavior as function of $f$. The $f$-dependencies of the reduced
frequency $\omega /2\pi $ and of the deviation $\Delta $ are shown in
Fig. 6.

Note that, representing the kink's law of motion in the form $\xi =\bar{v}%
t+\zeta (t)$, where $\zeta (t)$ is a time-periodic function with a small
amplitude, one can expand the kink solution as follows (here, the
"nonrelativistic" approximation $\bar{v}^{2}\ll 1$ is assumed):
\begin{eqnarray}
\phi (x,t)=4\tan ^{-1}\{\exp \left[ x-(\bar{v}t+\zeta (t))\right]
\}\nonumber\\
\approx
4\tan ^{-1}\{\exp (x-\bar{v}t)\}+2\zeta (t){\rm sech}(x-\bar{v}t).
\label{expansion}
\end{eqnarray}

The small oscillating term in Eq. (\ref{expansion}) may be a source for
emission of radiation, provided that it contains spectral components with
frequencies $\omega >1$ \cite{Kivshar:1989:RMP}. The Fourier
decomposition of the source,
defined as $\zeta (t)=\sum\limits_{n=-\infty }^{\infty }a_{n}\exp (in\omega
t)$, is illustrated by Fig. 7. Evidently, the density of the spectral
components decreases with the distance from the boundary between the pinning
and free-motion regions, i.e., with the increase of $f$. It is also obvious
that the intensity of the generated radiation is extremely small, as the
maximum frequency shown in 
%%%%%%%%%%%%%%%%%%%%%%%%%%%%%%%%%%%%%%%%%%%
\begin{figure}[t]
\centerline{\hbox{
\psfig{figure=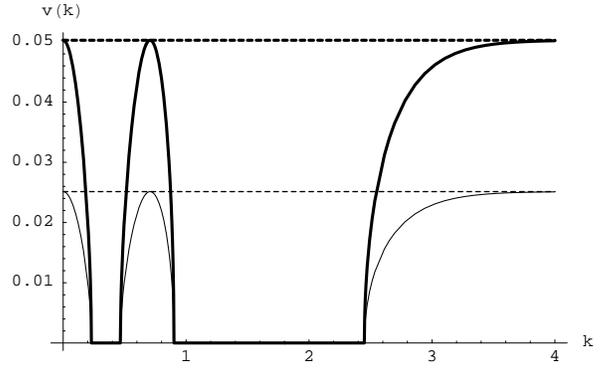,width=80mm,angle=0}}}
\caption{The average velocity $\overline{v}$ vs. the modulation wave number $%
k $ for $f=0.04$, $\eta =1$.}
\end{figure}
%%%%%%%%%%%%%%%%%%%%%%%%%%%%%%%%%%%%%%%%%%%
Fig. 7 is $\approx 0.377$ of the threshold value, 
$\omega =1$, beyond which the emission of radiation takes place.

(iii) The pinning regime, $\left| A/B\right| >1$. In Fig. 5, one has the
pinning ($\bar{v}=0$) and free-motion ($\overline{v}\neq 0$) regions of $k$
in accord with the diagram in Fig. 2. In the pinning regime, Eq. (\ref{12})
exhibits two sets of fixed points: potential minima, $\xi _{{\rm a}}=(\pi
/k)n-(1/2k)\sin ^{-1}\left( B/A\right) ,\ n=0,\,\pm 1,...$, which are
attractors (stable fixed points), and potential maxima, which are repellers
(unstable points), $\xi _{{\rm r}}=(4n-1)\pi /4k+(1/2k)\sin ^{-1}\left(
B/A\right) $. In the pinning regime, a time-dependent solution to Eq. (\ref
{12}) is (cf. Eq. (\ref{14}))\\
$\tan \left( k\xi (t)\right)$
\begin{equation}
 =-A/B+\sqrt{A^{2}/B^{2}-1}\tan \left[ k\sqrt{
A^{2}-B^{2}}(t-t_{0})\right] .
\label{15}
\end{equation}
This solution describes overdamped motion of a particle trapped in a well of
the tilted potential $\langle U_{{\rm eff}}\rangle $.
The stationary state is reached after a short transient.

\section{Numerical simulations}

\subsection{Direct simulations of the perturbed sine-Gordon equation
}

To get direct insight into the dynamics of the present weakly disordered
model and check the above analytical results, we solved Eq. (\ref{pert})
numerically. For this purpose, 
%%%%%%%%%%%%%%%%%%%%%%%%%%%%%%%%%%%%%%%%%%%
\begin{figure}[b]
\centerline{\hbox{
\psfig{figure=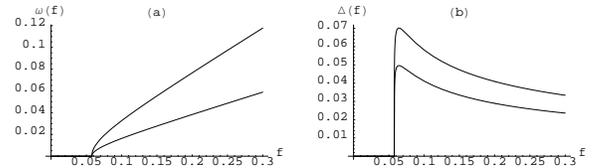,width=80mm,angle=0}}}
\caption{Frequency of the oscillating part of the velocity and the mean
square deviation of the trajectory from the linear one in the
free-motion regime (ii), $\omega (f)$ (a) and $\Delta (f)$
(b) for $\eta =1$, $k=1$.}
\end{figure}
%%%%%%%%%%%%%%%%%%%%%%%%%%%%%%%%%%%%%%%%%%%
%%%%%%%%%%%%%%%%%%%%%%%%%%%%%%%%%%%%%%%%%%%
\begin{figure}[hbt]
\centerline{\hbox{
\psfig{figure=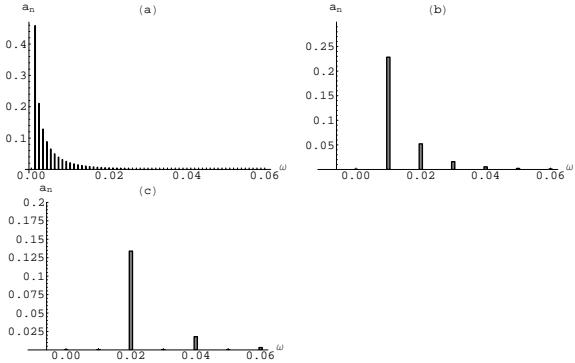,width=80mm,angle=0}}}
\caption{Fourier spectra of the periodic part of the trajectory for $\eta =1$%
, $k=1$, and $\Gamma =10$: $f=0.057942$ (a), $f=0.076369$ (b), and $%
f=0.115665$ (c).}
\end{figure}
%%%%%%%%%%%%%%%%%%%%%%%%%%%%%%%%%%%%%%%%%%%
Eq. (\ref{pert}) was transformed to a set
of difference equations in the spatial coordinate (differential in
the time coordinate)  and solved by using the
eighth-order explicit Runge-Kutta scheme \cite{Dormand:1980:JCAM},
with a stepsize control such that the time step was dynamically changed
within the range from $0.05$ to $0.3$. It has been chosen so that the
relative (per one site) error at each step did not exceeded $10^{-10}$.
 In order to study the kink dynamics within rather big time intervals (for our
computations the required time lap was $0 \leq t \leq 10^4$) we have
used a shifting computational domain with  boundaries which absorbed the
outgoing waves.

At $t=0$, we took, for the initial conditions, the well-known
approximate (perturbative) static kink solution valid in the
absence of the spatial disorder \cite{Kivshar:1989:RMP},
\begin{eqnarray}
\Phi _{K}(s) &=& \phi_s+4\tan ^{-1}(\exp s)\,,
\ s=\frac{x-v_0t}{\sqrt{1-v_0^2}} \, ,
\label{16} \\
v_0^2 &=& (1+(\Gamma/2\pi f)^2)^{-1}\ ,
\nonumber
\end{eqnarray}
where
\begin{equation}
\phi_s\equiv -\sin^{-1}f
\label{background}
\end{equation}
takes into account the shift of the background under the action of the
dc drive.

In the simulations we used $1000$ different realizations of the disorder
$\{\epsilon\}$ and then
averaged the results over the realizations with equal weights.

In Fig. 8, we compare the kink velocity, as obtained from the direct
simulations of Eq. (\ref{pert}), and those predicted by Eq. (\ref{12}).
The mean field velocity overestimates the exact one
what is a typical mean field feature.

The numerically found time dependence of the kink's velocity in the
localization regime is plotted in
Figs. 9 and 10. The corresponding dependence of the inverse total distance
travelled by the kink, in this regime, during the simulation time is
shown, as a function of the driving force, in Fig. 11.
It is evident from this figure  that, while with small $f$ the kink
quickly gets trapped,
 with the increasing $f$  it keeps moving for a much longer time.
%%%%%%%%%%%%%%%%%%%%%%%%%%%%%%%%%%%%%%%%%%%
\begin{figure}[hbt]
\centerline{\hbox{
\psfig{figure=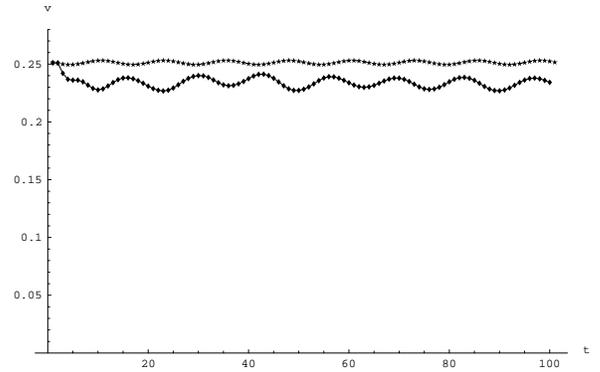,width=80mm,angle=0}}}
\caption{Comparison of the velocity-vs.-time curves obtained from the
numerically exact (diamonds) and adiabatic (stars) (Eq. (\ref{12}))
solutions for $ f=0.09,\ \eta =0.1,\ k=1$,\ and $\Gamma =2$.}
\end{figure}
%%%%%%%%%%%%%%%%%%%%%%%%%%%%%%%%%%%%%%%%%%%

 When approaching the critical point, the trapping time enormously
 increases, therefore the direct numerical
determination of the critical force $f_{\rm cr}$ from the dynamical
simulations is difficult. Besides, the calculations are constrained
to a finite size of the system, hence the critical field would be size
dependent.  Moreover, at low $\Gamma $, when $f/\Gamma >1$, the
one-kink model fails, as nucleation of kink-antikink pairs sets in
with the accumulation of energy. Therefore, investigation of the
critical region (transition between the trapped and freely moving kinks
is not really possible on the basis of the direct simulations
from Eq. (\ref{pert}).

\subsection{Numerical simulations of the collective coordinate equations
}

Assuming low fields $f$, we will now follow the Rice$^{'}$s
collective coordinate approach \cite{Rice:1983:PRB}, based on a modified ansatz for the kink,
cf. Eqs. (\ref{kink}) and (\ref{16})
\begin{equation}
\phi_K (x,t)= \phi_s+ 4\tan^{-1} \exp \left (\frac{x-\xi(t)}{L(t)}\right
)\,.
\label{21}
\end{equation}
The difference from Eq. (\ref{kink}) is that the present ansatz admits a
variable kink's width $L(t)$.
Inserting (\ref{21}) into the model's Hamiltonian
\begin{eqnarray}
H= \int\limits_{-\infty}^{\infty} dx \left \{\left( \Phi_t^2
+\Phi_x^2 \right )/2+ [1+ \epsilon (x)\sin (kx)]\right.\nonumber\\
\left.\times(\cos\phi_s-\cos\Phi)-f x\Phi_x\right \}, \hspace{2cm}
\label{22}
\end{eqnarray}
one arrives at the effective Hamiltonian \cite{Mingaleev:1999:PRB}
\begin{equation}
H_{\rm eff} = \frac{L}{16} p_{\xi}^2 +\frac{3 L}{4\pi^2}p_L^2 +U(L)+
V(\{\epsilon\}, L,\xi)+ 2\pi f \xi
\label{23}
\end{equation}
for the collective coordinates $\xi$, $L$ and the  canonically conjugate
momenta
%%%%%%%%%%%%%%%%%%%%%%%%%%%%%%%%%%%%%%%%%%%
\begin{figure}[t]
\centerline{\hbox{
\psfig{figure=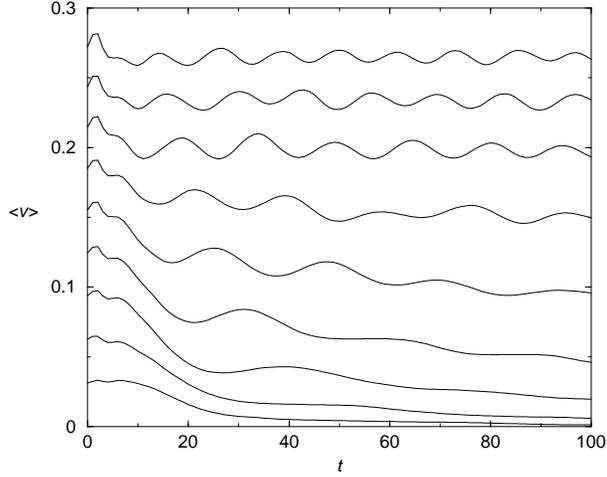,width=80mm,angle=0}}}
\caption{Evolution of the soliton velocity for different values of
the driving $f =  0.1, 0.2, 0.3, 0.4, 0.5, 0.6, 0.7$.
$\Gamma=2$, $k=1$, $\eta=0.1$ and initial position $x_0=12.5$.}
\end{figure}
%%%%%%%%%%%%%%%%%%%%%%%%%%%%%%%%%%%%%%%%%%%
\begin{equation}
p_{\xi}=\frac{8}{L}\frac{d \xi}{dt}, \qquad p_L =
\frac{2\pi^2}{3L}\frac{dL}{dt}.
\label{24}
\end{equation}

In the effective Hamiltonian (\ref{23}),
\begin{equation}
U(L)= 4L^{-1}+4\cos (\phi_s) L
\label{25}
\end{equation}
is the effective potential in the case without disorder, and
\begin{equation}
V(\{\epsilon\}, L, \xi)=  \int\limits_{-\infty}^{\infty} dx\,
\epsilon(x)\sin(kx)(\cos\phi_s-\cos\Phi)
\label{26}
\end{equation}
is an effective random potential.

Taking into account the friction, we obtain the following system of
equations of motion for the two kink's degrees of freedom:
\begin{eqnarray}
\frac{dp_{\xi}}{dt} +\frac{\Gamma}{8}p_{\xi}+ \frac{d}{d\xi} V(\{\epsilon\}, L,
\xi)+2\pi f= 0, \qquad
\label{27}\\
\frac{dp_L}{dt} +\frac{\Gamma}{8}p_L+
\frac{3p7_L^2}{4\pi^2}+\frac{p_{\xi}^2}{16} + \frac{d}{dL}[U(L)+
V(\{\epsilon\}, L, \xi)]= 0.
\label{28}
\end{eqnarray}

Figs. 12 and 13 display comparison of the kink's
velocities and widths, as obtained from the direct simulations of Eq.
(\ref{pert}) and from simulations of the collective coordinate equations
(\ref{27}) and (\ref{28}). A satisfactory agreement between the results of
the two approaches is evident.

Eq.(\ref{27}) makes it possible to determine the critical force
$f_{\rm cr}$ necessary to detrap the kink.
For a single potential well (trap), $f_{\rm cr}$ is simply equal to the
maximum value of the local force induced by the trapping
potential. However, a spatially random potential assumes the presence of
a collection of the wells with randomly distributed parameters.
Strictly speaking, one cannot uniquely define a global value
of $f_{\rm cr}$ in this case. Instead, it is necessary to consider a
dynamical problem of gradually detrapping the kinks with the increase of
the driving force \cite{Malomed:1989:PRB}.
%%%%%%%%%%%%%%%%%%%%%%%%%%%%%%%%%%%%%%%%%%%
\begin{figure}[t]
\centerline{\hbox{
\psfig{figure=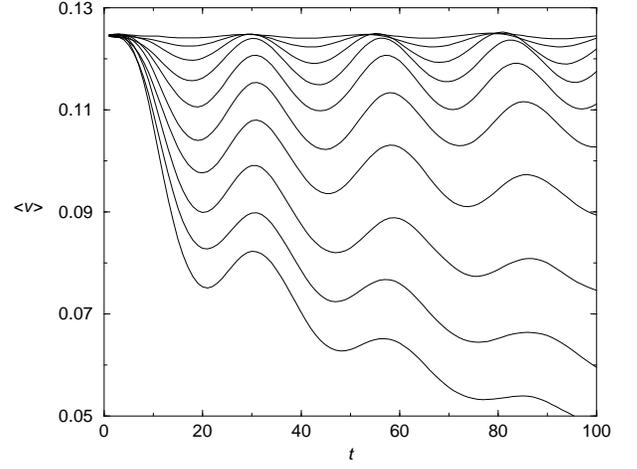,width=80mm,angle=0}}}
\caption{The same as in Fig. 9, for $\eta=0.01,..., 0.1 $, $f=0.04$,
$\Gamma=2$, $k=1$, and initial position $x_0=12.5$ for $5.10^3$
realizations of the random potential.}
\label{}
\end{figure}
%%%%%%%%%%%%%%%%%%%%%%%%%%%%%%%%%%%%%%%%%%%%%
Nevertheless, it is possible to give a reasonable definition of the
critical detrapping force as that which is equal to the local maximum
value of the spatially random force, {\em averaged} over the length of
the system (i.e., over all the local maxima):
\begin{equation}
2\pi f_{\rm cr}= \left\langle\frac{d}{d \xi} V(\{\epsilon\}, \xi
)\right\rangle_{\rm max},
\label{fcrit}
\end{equation}
the angular brackets standing for averaging.
Here, the kink's width in the potential $V$ (\ref{26}) should be set
equal to the unperturbed value $L_0=1$.

If we suppose equal statistical weights for various local extrema  of
the function $dV/d\xi$, then the averaging theorem from Ref.
\cite{Kree:1986:MRP}
(used e.g. in \cite{Mingaleev:1999:PRB}) can be applied directly, yielding

\begin{equation}
\left\langle\frac{d}{d \xi} V(\{\epsilon\}, X)\right\rangle_{\rm max}= \sqrt
\frac{\pi}{2}\frac{M_2}{\sqrt M_4},
\label{M2M4}
\end{equation}
%%%%%%%%%%%%%%%%%%%%%%%%%%%%%%%%%%%%%%%%%%%
\begin{figure}[hbt]
\centerline{\hbox{
\psfig{figure=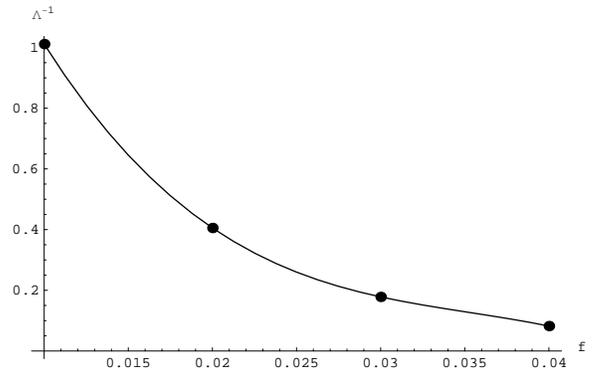,width=80mm,angle=0}}}
\caption{Dependence of the inverse mean trajectory $\Lambda^{-1}$
on the force $f$ for $\Gamma=2$, $k=1$, $\eta=0.1$ and
initial position $x_0=12.5$.}
\end{figure}
%%%%%%%%%%%%%%%%%%%%%%%%%%%%%%%%%%%%%%%%%%%
%%%%%%%%%%%%%%%%%%%%%%%%%%%%%%%%%%%%%%%%%%%
\begin{figure}[th]
\centerline{\hbox{
\psfig{figure=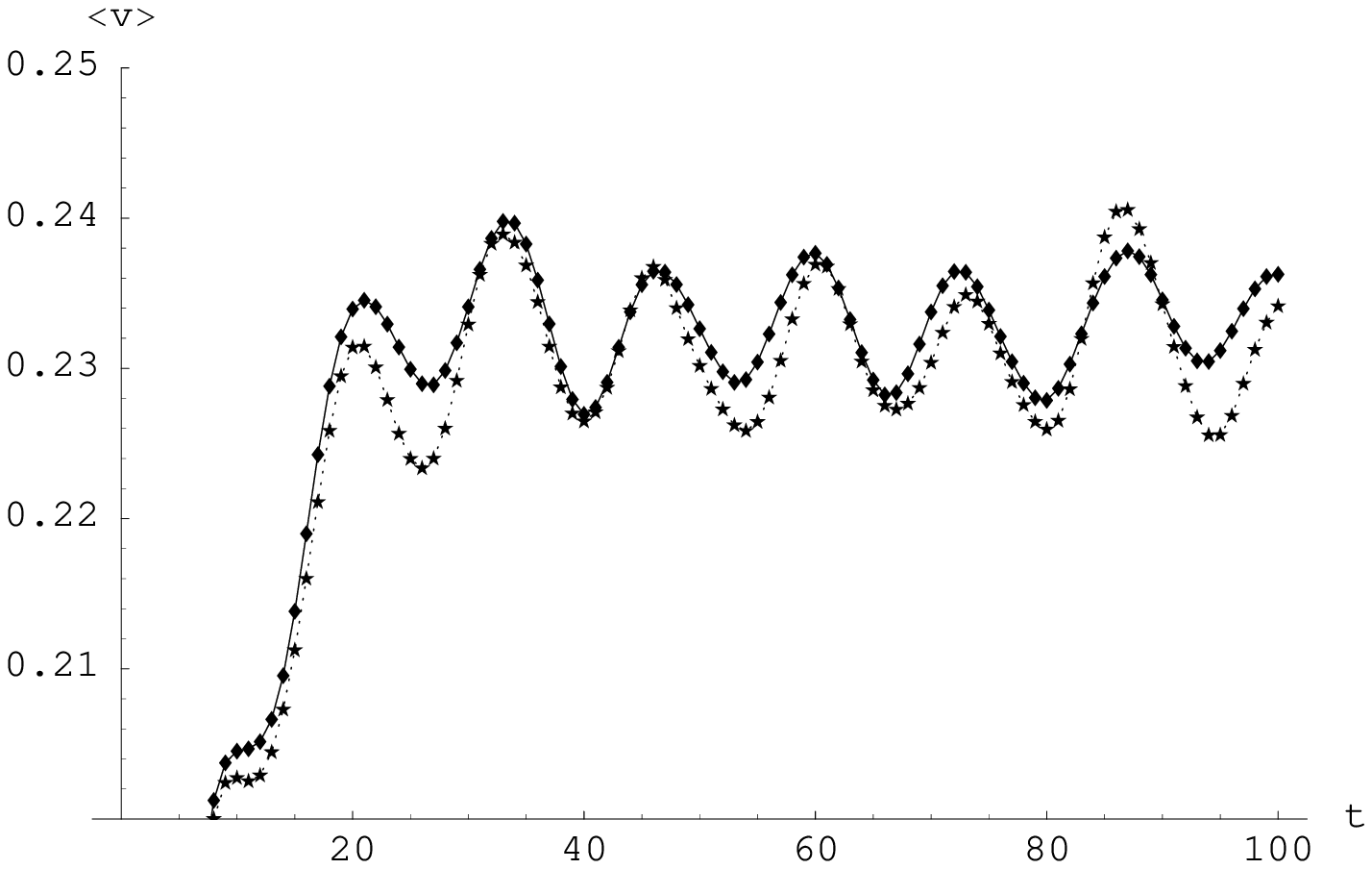,width=80mm,angle=0}}}
\caption{Comparison of the averaged velocities vs. time from the direct
numerical simulations of Eq. (\ref{pert}) (squares) with that from the
numerical solution of the collective coordinate equations (diamonds) for
$\eta= 0.1, \, f=0.08, \, k=1, \,\Gamma=2. $} 
%\caption{ The same for the averaged widths at the same values of
%parameters as in Fig. 13.}
\end{figure}
%%%%%%%%%%%%%%%%%%%%%%%%%%%%%%%%%%%%%%%%%%%

where, in the present case,
\begin{equation}
M_2=\left\langle \left ( \frac{d^2 V}{d \xi^2}\right ) ^2\right \rangle =
4\eta^2 \int\limits_{-\infty}^{\infty} d x \sin^2(k x)q^{''}
(z),
\label{M2}
\end{equation}
\begin{equation}
M_4=\left\langle \left ( \frac{d^3 V}{d \xi^3}\right ) ^2\right\rangle =
4\eta^2 \int\limits_{-\infty}^{\infty}d x \sin^2(k x)q^{'''}
(z),
\label{M4}
\end{equation}
and $2q(z)\equiv \left(\cos\phi_s\right ){\rm sech}^2 (z)-\left
(\sin\phi_s\right) \sinh (z){\rm sech}^2(z)$ (with $z\equiv (x-\xi)/L$).

The integrals in (\ref{M2}) and (\ref{M4}) can be evaluated in a
straightforward way,
\begin{equation}
I_j= (1-f^2_{\rm cr})Q_j+f^2_{\rm cr} R_j, \quad j=1,2;
\label{I}
\end{equation}
where
\begin{equation}
Q_1  = \frac{32}{21}-16\cos (2k\xi)S(\pi k L)\left (
\frac{2}{3}P_1-\frac{2}{5}P_2 +\frac{1}{35}P_3\right ) ,
\end{equation}
\begin{equation}
R_1  =   \frac{31}{21}-\cos (2k\xi)S(\pi kL)
\left ( 1-\frac{26}{3}P_1+\frac{32}{5}P_2-\frac{16}{35}P_3\right ) ,
\end{equation}
\begin{eqnarray}
Q_2 =  \frac{256}{15}-128\cos (2k\xi)S(\pi kL)
 \left ( \frac{2}{3}P_1-\frac{14}{15}P_2\right.\nonumber\\
 \left.+\frac{4}{21}P_3-\frac{2}{315} P_4\right ),
\end{eqnarray}
\begin{eqnarray}
R_2  =  \frac{254}{15}-\cos (2k\xi)S(\pi KL)
\left ( 1-\frac{80}{3}P_1+\frac{896}{15}P_2\right.\nonumber\\
\left.-\frac{256}{21} P_3 +\frac{2.24^2}{9^2.35}P_4\right ),
\label{Ints}
\end{eqnarray}

%%%%%%%%%%%%%%%%%%%%%%%%%%%%%%%%%%%%%%%%%%%
\begin{figure}[th]
\centerline{\hbox{
\psfig{figure=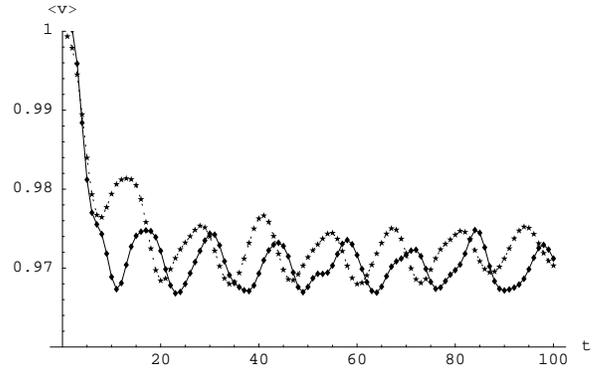,width=80mm,angle=0}}}
%\caption{Comparison of the averaged velocities vs. time from the direct
%numerical simulations of Eq. (\ref{pert}) (squares) with that from the
%numerical solution of the collective coordinate equations (diamonds) for
%$\eta= 0.1, \, f=0.08, \, k=1, \,\Gamma=2. $} 
\caption{The same for the averaged widths at the same values of
parameters as in Fig. 13.}
\end{figure}
%%%%%%%%%%%%%%%%%%%%%%%%%%%%%%%%%%%%%%%%%%%
where $S(x)=x/\sinh x\ , P_n=\prod\limits _{l=1}^n ((kL)^2+l^2)$.
Then, for $f_{\rm cr}$, one arrives at expression
\begin{equation}
f_{cr}= \frac{\eta}{\sqrt{2\pi L}}\frac{I_1(f^2_{\rm cr})}{\sqrt
{I_2(f^2_{\rm cr}})}.
\label{fcrit1}
\end{equation}

The resulting function $f_{\rm cr}(kL,\xi)$  is plotted in Fig. 14.
To compare it with the mean-field result shown above in Fig. 2, we have
set here $\xi=0$.
 A drastic difference between the two results is evident.
 Note also that the linear dependence of $f_{\rm cr}$ on the disorder
 strength $\eta$ is in contrast to the quadratic dependence on $\eta $
 in the mean field result (\ref{border}).

It is relevant to stress that a characteristic critical depinning force,
which was found for the model (\ref{SG}) with the unmodulated disorder
in Ref. \cite{Malomed:1989:PRB}, also scaled linearly with $\eta$. In other words,
the depinning process is dominated by the underlying spatial disorder,
rather than by the fact that the disorder is subject to a regular
modulation. Thus, the mean-field approximation (alias, straightforward
perturbation theory for the weak disorder) can not adequately
describe depinning. Nevertheless, the modulation gives rise to a
dependence of the pinning threshold upon the modulation period, see
Fig. 14. In the limit of large $k$, the oscillations are dominated by
the randomness and the case is reduced to that of the pure random
noise. However, the most interesting is the region of the length scale
competition $k^{-1} \sim L\leq 1$.

Note, that a strong nonperturbative response of solitons
to weak random fluctuations in a narrow
vicinity of a transition between different states
is also known in other models, e.g., a dual-core nonlinear
optical fiber \cite{Mostofi:1998:OC}.

\section{Conclusion}
In this work, we have introduced a dc-driven damped sine-Gordon model
with a small
nonlinear spatial-disorder term, on which a sinusoidal modulation is
superimposed. It describes, e.g., a weakly disordered system with a regular
grain structure. It was demonstrated that, in the mean-field
approximation (second order of the perturbation theory),
the periodically modulated disorder gives rise to
an effective periodic potential. Dynamics of a kink moving in this potential
was studied in the overdamped limit, using the adiabatic approximation,
the main objective being to consider depinning of a trapped kink.
The analytical results were compared
against direct numerical simulations of the underlying model, as well as
against numerical results using the collective-coordinate approach but
without the mean-field approximation. It has been found that depinning
of a kink trapped by the weak spatial disorder is, in reality,
drastically different from what predicted by the mean-field
approximation.
%%%%%%%%%%%%%%%%%%%%%%%%%%%%%%%%%%%%%%%%%%%
\begin{figure}[th]
\centerline{\hbox{
\psfig{figure=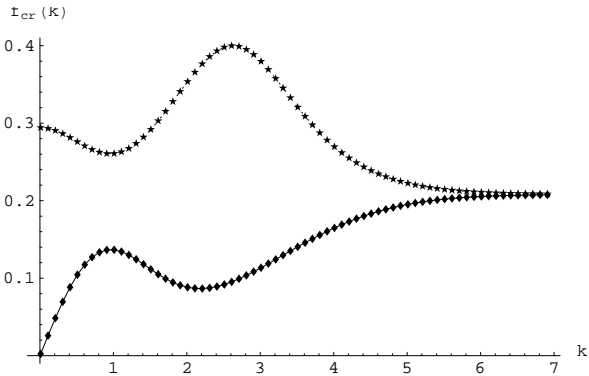,width=80mm,angle=0}}}
\caption{Critical field $f_{\rm cr}$ vs. wave number of the modulating field $k$.
Full line is for $\xi=\pi/2k$, dashed for $\xi=0$.
Quantitative as well as the qualitative difference to the mean field
version in Fig. 2 is evident.} 
\end{figure}
%%%%%%%%%%%%%%%%%%%%%%%%%%%%%%%%%%%%%%%%%%%

\section*{Acknowledgement}
We thank S.F. Mingaleev for discussions on numerical simulation methods.
B.A.M. appreciates hospitality of the Department of Theoretical Physics at
the Palack\'{y} University (Olomouc, The Czech Republic). The support by the
grant No. 202/97/0166 of the Grant Agency of the Czech Republic and
partially also by the grant No. 2/4109/98 of the VEGA Grant Agency is
greatly acknowledged.

%%%%%%%%%%%%%%%%%%%%%%%%%%%%%%%%%%%%%%%%%%%%%%%%%%%%%%%%%%%%%
%\bibliographystyle{ieeetr}
%\bibliographystyle{prsty}
%\bibliography{abbrev,kg,lri-sol,gen-sol,sg-nonloc}
%%%%%%%%%%%%%%%%%%%%%%%%%%%%%%%%%%%%%%%%%%%%%%%%%%%%%%%%%%%%%

\end{document}